\documentclass[prl,aps,twocolumn,showpacs,preprintnumbers,amsmath,amssymb]{revtex4}
\usepackage{epsfig}
\usepackage{color}
\usepackage{ifthen}
\usepackage{xfrac}
\usepackage{graphicx}
\usepackage{amsmath}
\usepackage{amssymb}

\usepackage{graphicx}
\usepackage{dcolumn}
\usepackage{bm}
\def\Dsl{\hbox{/\kern-.6700em\it D}} 
\def\dsl{\hbox{/\kern-.5300em$\partial$}}

\def\eqa{\begin{eqnarray}}
\def\eeqa{\end{eqnarray}}
\def\eq{\begin{equation}}
\def\eeq{\end{equation}}
\def\be{\begin{equation}}
\def\ee{\end{equation}}
\def\bea{\begin{eqnarray}}
\def\eea{\end{eqnarray}}

\newcommand{\dslash}{\not{\hbox{\kern-2pt $\partial$}}}
\newcommand{\pslash}{\not{\hbox{\kern-2.3pt $p$}}}
 \newtoks\nslashfraction
 \nslashfraction={.13}
 \newcommand{\nslash}[1]{\setbox0\hbox{$ #1 $}
   \setbox0\hbox to \the\nslashfraction\wd0{\hss \box0}/\box0 }

\begin{document}

\preprint{}

\title{Integrating out Heavy Fields in Inflation}
\author{Mark G. Jackson}
\affiliation{Paris Centre for Cosmological Physics and Laboratoire AstroParticule et Cosmologie, Universit\'{e} Paris 7-Denis Diderot \\ B\^{a}timent Condorcet, 10 rue Alice Domon et L\'{e}onie Duquet, 75205 Paris, France}

\date{\today}


\begin{abstract}
\noindent
I present the procedure for integrating out quantum fields whose mass $M$ is well above the Hubble scale $H$ in de Sitter space.  The effective interaction and density matrix are explicitly computed for a simple example and are found to be of order $H/M$, are nonlocally distributed in time and contain a highly oscillatory phase.  This radical departure from the static space-time result demonstrates that these new tools are necessary for constructing effective inflationary potentials.  The effective coupling naturally forms a perturbative series in the relative energy transferred $\Delta E/M$, whose coefficients can then be compared to observation.
\end{abstract}

\maketitle

\subsection{Introduction}
One of the major endeavors of modern theoretical physics is to understand quantum field theory in time-dependent backgrounds \cite{Birrell:1982ix}.  While long of formal interest, this is now of practical importance given current and future cosmological data \cite{planck} which can discern between competing models of inflation \cite{Guth:1980zm, Linde:1981mu, Albrecht:1982wi, Linde:1983gd} and possibly even quantum gravity such as superstring theory.

Recent developments \cite{Jackson:2010cw, Jackson:2011qg, Jackson:2012fu} have allowed the computation of primordial correlation function corrections from high-energy physics in quasi-de Sitter space.  Although these are of interest for making connection to the cosmological data, a more fundamental question is how to construct the low-energy effective action of such a high-energy model.  This effective action can then be used for \emph{any} purpose, without knowledge of the original high-energy action.  In flat space-time this construction is trivial: the Green's function for a heavy field can be expanded around the low-energy limit,
\begin{eqnarray}
\nonumber
\mathcal G(x,y) &=&  \int \frac{d^4p}{(2 \pi)^4} \frac{e^{-ip \cdot (x-y)}}{p^2 - M^2} \\
\label{gfnaive}
&\approx& - \left( \frac{1}{M^2} - \frac{\partial^2}{M^4} + \cdots \right)  \delta^4(x-y).
\end{eqnarray}
This effective interaction is local and its magnitude of order $1/M^2$.  An expanding background (such as quasi-de Sitter space occurring during inflation), on the other hand, mixes the energy scales in such a way that a new set of tools is called for \cite{Brandenberger:1999sw}.  Previous studies \cite{Schalm:2004qk,Easther:2001fi, Easther:2001fz, Easther:2002xe, Kaloper:2002uj, Kaloper:2002cs, Burgess:2002ub} have attempted to address this question but did not allow for energy non-conservation or a non-trivial density matrix.

In this article we take the first step in this direction.  We begin with the action of a light field coupled to a heavy one in de Sitter space-time.  The heavy field is then integrated out using the techniques developed in \cite{Jackson:2010cw, Jackson:2011qg, Jackson:2012fu}, yielding an effective potential and density matrix for the light field.  The result will differ markedly from the flat space-time result of (\ref{gfnaive}).
\subsection{Effective Action}
We will consider the theory containing light field fluctuations $\varphi$ coupled to a heavy field $\chi$:
\begin{equation}
\label{uvaction}
S = - \int d^4 x \sqrt{g} \left[ \frac{1}{2} (\partial \varphi)^2 +  \frac{1}{2} (\partial \chi)^2 +  \frac{M^2}{2} {\chi}^2 +  \frac{g}{2} \varphi^2 \chi \right] . 
\end{equation}
The UV vacuum is assumed to be Bunch-Davies \cite{Bunch:1978yq}.  The fact that this is an unstable potential need not worry us as we are only concerned with the perturbative interactions to illustrate the effective action construction.

\begin{figure}
\begin{center}
\hspace{-0.0in}
\parbox{25mm}{\includegraphics[scale=0.18]{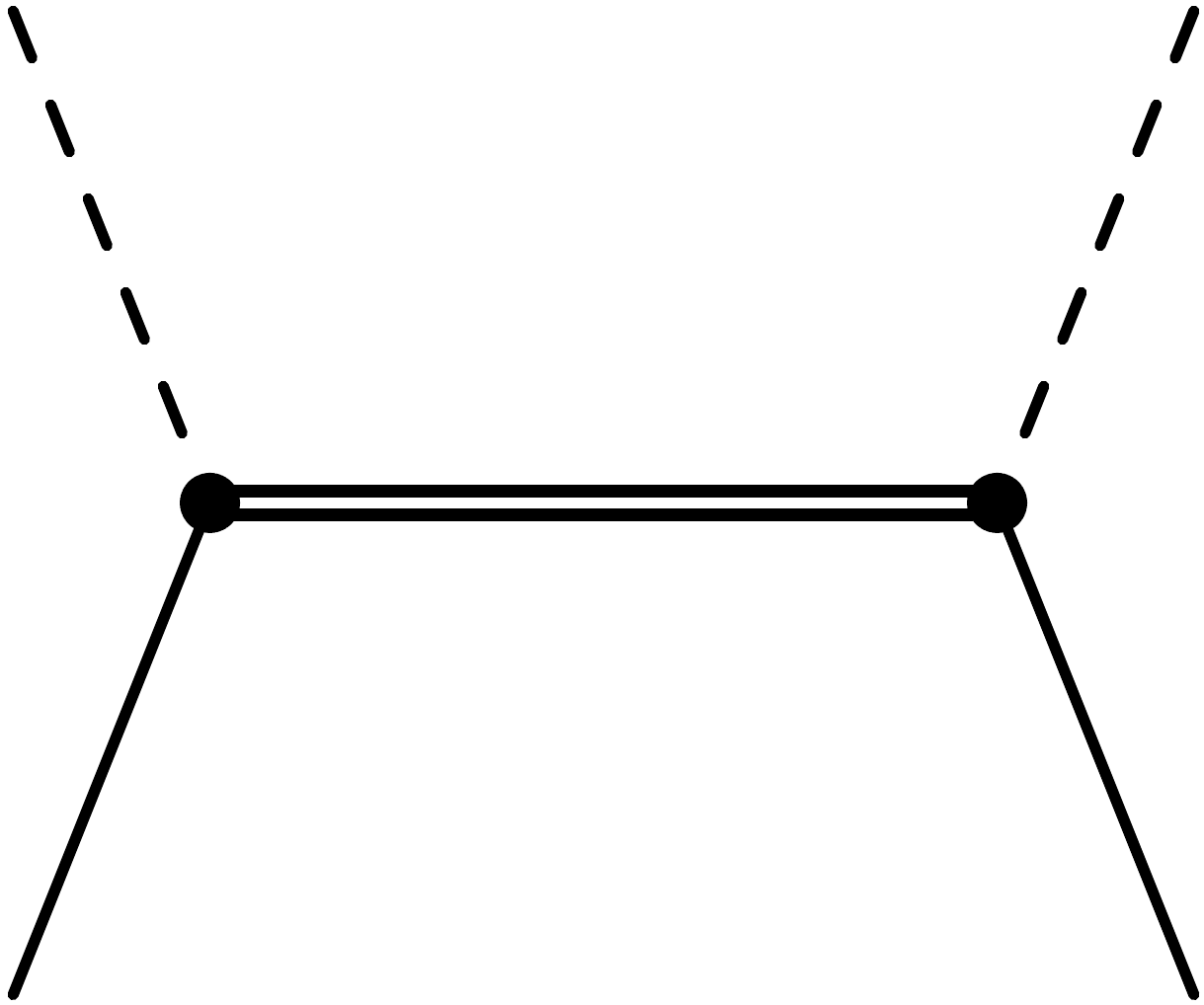} \hspace{-0.1in} \newline A} \ \
\parbox{25mm}{\includegraphics[scale=0.18]{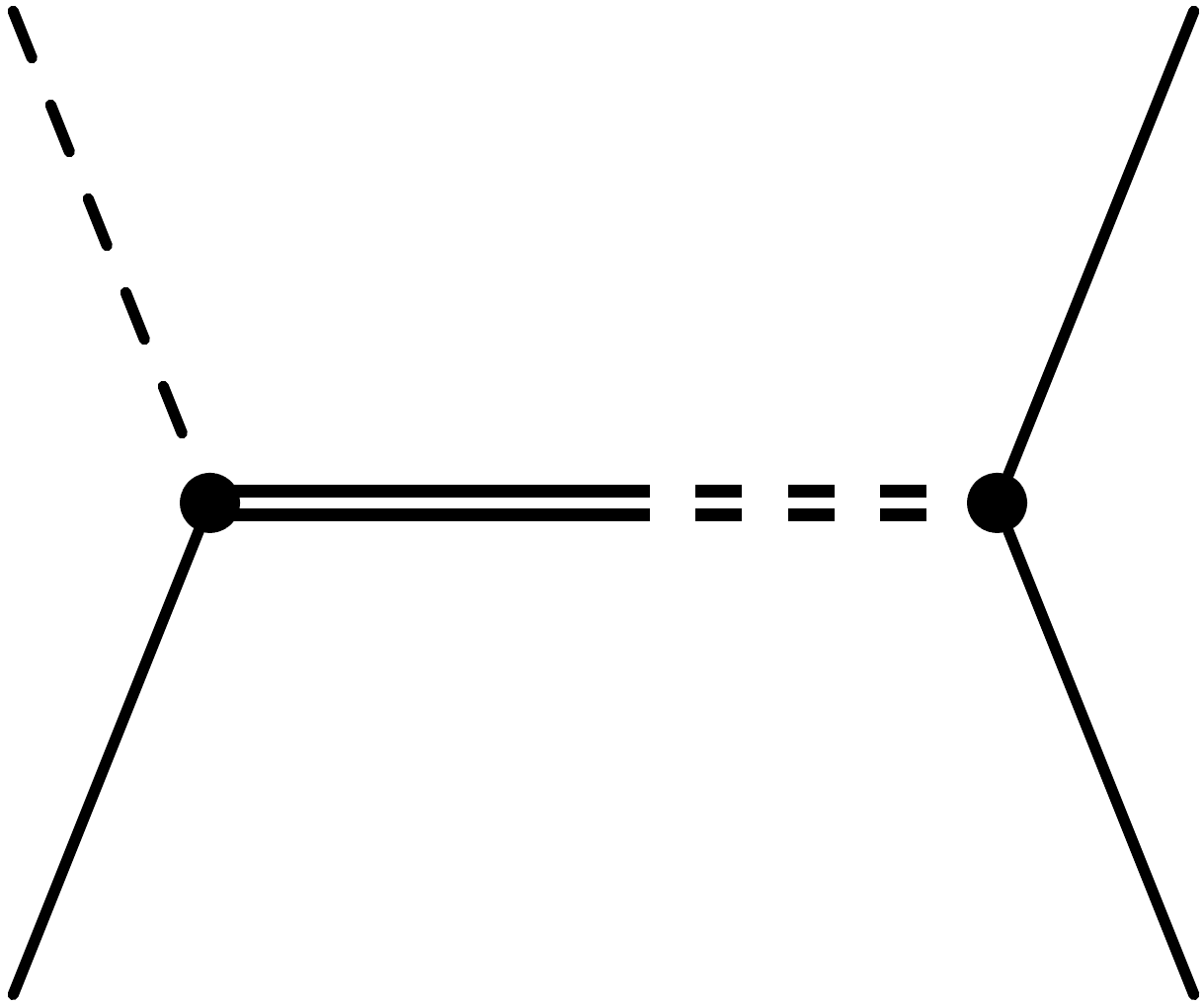} \hspace{-0.1in} \newline B} \ \
\parbox{25mm}{\includegraphics[scale=0.18]{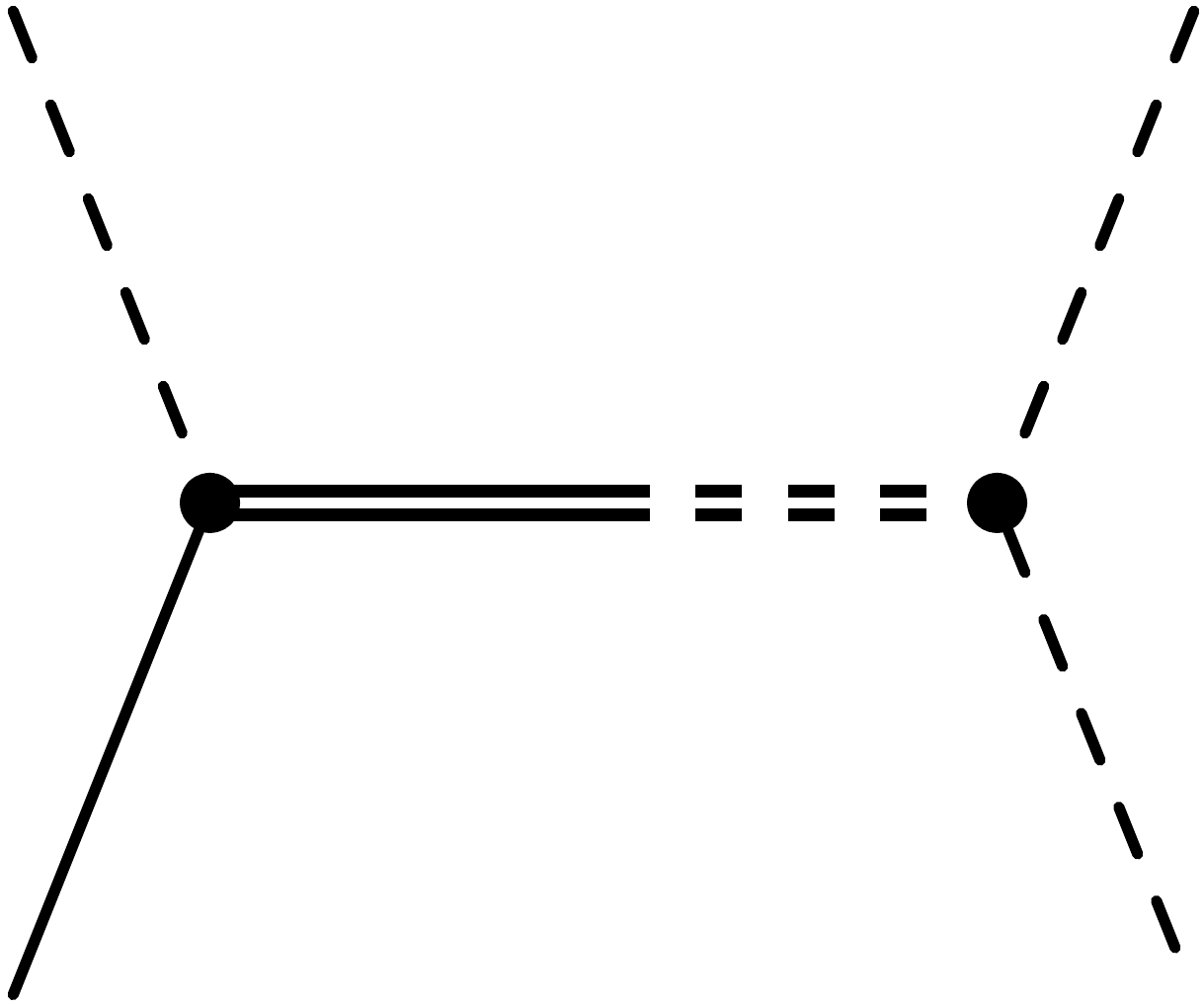} \hspace{-0.1in} \newline C} \\
\caption{Integrating out the $\chi$-field induces interactions between the $\varphi$-fields.  Single solid lines indicate  ${\bar \varphi}$, dashed single lines indicate $\Phi$.  The double lines indicate the analogous heavy field components $\{ {\bar \chi},{\rm X} \}$.}
\label{eff}
\end{center}
\end{figure}

The Schwinger-Keldysh (or ``in-in") formalism is the appropriate one for non-equilibrium systems \cite{Calzetta:1986cq}.  Define 
\begin{eqnarray*}
&& {\bar \varphi} \equiv (\varphi_+ + \varphi_-)/2, \hspace{0.2in} \Phi \equiv \varphi_+ - \varphi_-, \\
&& {\bar \chi} \equiv (\chi_+ + \chi_-)/2, \hspace{0.2in} {\rm X} \equiv {\chi}_+ - {\chi}_-
 \end{eqnarray*}
so that the action $\mathcal S \equiv S[\varphi_+, \chi_+] - S[\varphi_-, \chi_-]$ equals
\begin{eqnarray}
\nonumber
&& \hspace{-0.2in} \mathcal S [ {\bar \varphi}, {\Phi}, {\bar \chi},  {\rm X}] =  - \int d^4 x \sqrt{g} \Big[ \partial {\bar \varphi} \partial \Phi + \partial {\bar \chi} \partial {\rm X} + M^2  {\bar \chi} {\rm X} \\
\label{sorig}
&& \hspace{0.5in} \left. + g \left( {\bar \varphi} \Phi {\bar \chi} + \frac{1}{2} {\bar \varphi}^2 {\rm X} + \frac{1}{8} {\Phi}^2 {\rm X} \right) \right].
\end{eqnarray}
Performing the path integral over the $\chi$-field yields the action ($\mathcal F \equiv \langle {\bar \chi} {\bar \chi} \rangle$ is the Wightman function and $\mathcal G^R \equiv~i \langle {\bar \chi} {\rm X} \rangle$ the retarded Green's function for the heavy field)
\begin{eqnarray}
\label{snochi}
&& \hspace{-0.2in} \mathcal S_{\rm eff}[ {\bar \varphi}, \Phi] = - \int d^4 x \sqrt{g} \partial {\bar \varphi} \partial \Phi \\
\nonumber
&& + \frac{g^2}{2} \int d^4 x_1 \sqrt{g} \int d^4 x_2 \sqrt{g} \Big[ i {\bar \varphi} \Phi(x_1) \mathcal F(x_1,x_2)  {\bar \varphi} \Phi(x_2) \\
\nonumber
&& \hspace{-0.0in} - \frac{1}{2} {\bar \varphi} \Phi(x_1) \mathcal G^R(x_1,x_2)  \left( {\bar \varphi} (x_2)^2 + \frac{1}{4}  \Phi(x_2)^2 \right) \Big].
\end{eqnarray}
These interactions are shown in Figure~\ref{eff}.  The $i$ in the first interaction indicates a correction to the density matrix, not the dynamical part of the action.  This theory now contains only $\varphi$-modes, but of varying energies.  The remainder of this article derives the approximation of $\mathcal F$ and $\mathcal G^R$ to leading order in $H/M$.

\subsection{Diagram A}
By transforming into comoving momentum-space and comoving time $\tau$, the interaction term in (\ref{snochi}) corresponding to diagram A can be written 
\begin{eqnarray}
\label{sma}
\mathcal S^A_{\rm eff} [ {\bar \varphi}, \Phi ] &=&\frac{i g^2}{4} \int d \tau_1\ a(\tau_1)^4 \ d \tau_2 \ a(\tau_2)^4 \\
\nonumber
&&  \hspace{-0.4in} \times {\bar \varphi}_{{\bf k}_1}  {\Phi}_{{\bf k}_2} (\tau_1) \left[ V_{{\bf k}_{12} } (\tau_1) V^*_{{\bf k}_{12} } (\tau_2) \right. \\
\nonumber
&& \left. +  V^*_{{\bf k}_{12} } (\tau_1) V_{{\bf k}_{12} } (\tau_2) \right] {\bar \varphi}^*_{{\bf k}_3}  {\Phi}^*_{{\bf k}_4} (\tau_2)
\end{eqnarray}
where I have not written the integrals over ${\bf k}$ but assume that ${{\bf k}_{12} } \equiv {\bf k}_1 + {\bf k}_2 = {\bf k}_3 + {\bf k}_4$, and
\begin{equation}
\label{vdef}
V_{\bf k} (\tau) \approx \frac{1}{a(\tau)} \frac{\exp \left[ - i \int ^\tau_{\tau_{\rm in}} d \tau' \sqrt{  k^2 + \frac{M^2}{H^2 \tau^{\prime 2}}} \right]}{\sqrt{ 2} \left( k^2 + \frac{M^2}{H^2 \tau^2} \right)^{1/4} } 
\end{equation}
is the WKB approximation for the free field solution of the heavy field.

Now expand each ${\bar \varphi} {\Phi}$-pair (with the scale factor-squared removed) in its Fourier conjugate,
\begin{equation}
\label{ft}
{\bar \varphi}_{{\bf k}_1}  {\Phi}_{{\bf k}_2} (\tau) = \frac{1}{a(\tau)^2} \int \frac{d \omega}{2 \pi} e^{-i \omega \tau} {\tilde {\bar \varphi}_{{\bf k}_1}} {\tilde  {\Phi}_{{\bf k}_2}} (\omega). 
 \end{equation}
Note that $\omega$ can take positive and negative values, a crucial point in what follows.  While in more general cases we would Fourier expand the ${\bar \varphi},{\Phi}$ fields separately, in this case we will only be concerned with the sum of their frequencies and so we can consider the two fields together.

This expansion allows us to perform the integrals over interaction times.  Consider the first configuration,
 \begin{eqnarray}
 \nonumber
\mathcal C_1(\omega, {\bf k}) & \equiv& \int^0_{-\infty} d \tau \ a(\tau)^2 e^{-i \omega \tau} V^*_{{\bf k}}(\tau) \\
\label{c1}
 &=&  \frac{1}{H }  \int^0_{\tau_{\rm in}} \frac{d \tau}{\tau} \frac{1}{\sqrt{2} \left(k^2 + \frac{M^2}{H^2 \tau^{2}} \right)^{1/4} } \times  \\
  \nonumber
 && \exp -i \left[ \omega \tau - \int^\tau_{\tau_{\rm in}} d \tau' \sqrt{k^2 + \frac{M^2}{H^2 \tau^{'2}} } \right].
 \end{eqnarray}
As observed previously \cite{Jackson:2010cw} this can be rescaled via $u~\equiv~\frac{H}{M}~\tau$ which allows a stationary phase approximation for $\omega > k$ at the condition
\begin{equation}
\label{statphase}
\omega = \sqrt{ k^2 + u_c^{-2}}
\end{equation}
which is solved to yield the (rescaled) time of interaction
\[ u_c^{-1} = - \sqrt{\omega^2 - k^2} . \]
The vertex (\ref{c1}) can then be evaluated to leading order in $H/M$ to be
\begin{eqnarray*}
\mathcal C_1(\omega, {\bf k}) &\approx& \sqrt{ \frac{\pi i }{HM}}  \left( \omega^2 -k^2 \right)^{-1/4} e^{-i \frac{M}{H} \sqrt{ k^2  u_{\rm in}^2+ 1}}    \\
&&\hspace{-0.2in} \times  \left( \frac{ {\omega  + \sqrt{ \omega^2 -k^2 }}}{\sqrt{k^2 +  u_{\rm in}^{-2} } + |u_{\rm in}|^{-1}} \right) ^{i \frac{M}{H} } .
\end{eqnarray*}

A similar vertex, but with the heavy field conjugated, can be evaluated in a similar way for $\omega < -k$.  This produces an identical answer but with opposite phase:
 \begin{eqnarray*}
 \nonumber
\mathcal C_2(\omega, {\bf k}) & \equiv& \int^0_{-\infty} d \tau \ a(\tau)^2 e^{-i \omega \tau} V_{{\bf k}}(\tau)  \\
&=& \mathcal C_1^* (- \omega, {\bf k}).
\end{eqnarray*}

The interaction term (\ref{sma}) can then be expressed in terms of these vertex functions as
\begin{eqnarray}
\label{sma0}
&& \hspace{-0.2in} \mathcal S^A_{\rm eff} = \frac{i g^2}{4} \int_{k_{12}}^\infty \frac{d \omega_1}{2 \pi} \frac{d \omega_2}{2 \pi} \left[ \mathcal C_1(\omega_1, {\bf k}_{12})  \mathcal C_1^*(\omega_2,{\bf k}_{12}) \right. \\
\nonumber
&& \left. + \mathcal C^*_1(-\omega_1, {\bf k}_{12})  \mathcal C_1(-\omega_2,{\bf k}_{12}) \right]  {\tilde {\bar \varphi}}_{{\bf k}_1}  {\tilde \Phi}_{{\bf k}_2} (\omega_1 ) {\tilde {\bar \varphi}}^*_{{\bf k}_3} {\tilde \Phi}^*_{{\bf k}_4} (\omega_2) .
\end{eqnarray}
Thus the two interactions occur at different times depending on the relative value of $\omega_1$ and $\omega_2$, producing a rapidly oscillating phase factor in $\mathcal C_1 \mathcal C^*_1$. Now defining the frequencies
\begin{equation}
\label{freqcombos}
 \Omega \equiv \omega_1 - \omega_2, \hspace{0.2in} {\bar \omega} \equiv \frac{1}{2} (\omega_1 + \omega_2), \\
\end{equation}
the oscillating part of $\mathcal C_1 \mathcal C^*_1$ can be approximated as
\begin{equation}
\label{osc}
 \left( \frac{ {\omega_1 + \sqrt{ \omega_1^2 - {k}_{12}^2 }}}{{\omega_2 + \sqrt{ \omega_2^2 - {k}_{12}^2 }}} \right)^{i \frac{M}{H} } \approx \exp \left( - i \frac{M}{H}\Omega {\bar u}_c \right) 
 \end{equation}
where 
\[  {\bar u}_c^{-1} \equiv - \sqrt{ {\bar \omega}^2 - {k}_{12}^2} \]
is the average inverse time between the two interactions~\cite{Jackson:2012fu}.  To understand what this approximation means, multiply the stationary phase condition (\ref{statphase}) by $H |\tau_c|$ allowing one to define the interaction energy as 
\[ E = H \omega | \tau_c| = \frac{M \omega}{\sqrt{\omega^2-k_{12}^2}}. \]
This provides an estimate of the energy transferred through the virtual heavy field, which we assume is small:
\[ \frac{\Delta E}{M} = \frac{\omega_1}{\sqrt{\omega_1^2-k_{12}^2}} - \frac{\omega_2}{\sqrt{\omega_2^2-k_{12}^2}} \sim \frac{\Omega}{\sqrt{{\bar \omega}^2-k_{12}^2}} \ll 1. \]
This is precisely what we would expect: if the energy transferred is of order $M$, we shouldn't try to construct an effective theory in which all modes above $M$ have been integrated out.  For now, we keep only the leading term in the expansion of $|\Omega {\bar u}_c| \ll 1$.  

Inverting the Fourier transform of (\ref{ft}), and using the rescaled coordinates $u \equiv \tau H/M$, $U \equiv T H/M$, the action becomes
\begin{eqnarray}
\nonumber
&& \hspace{-0.3in} \mathcal S^A_{\rm eff} = \frac{i \pi g^2}{4HM} \int_{k_{12}}^\infty \frac{d {\bar \omega}}{2 \pi}  \left( \int_0^{{\bar \omega}-k_{12}} \frac{d \Omega}{2 \pi} + \int_{-({\bar \omega}-k_{12})}^0 \frac{d \Omega}{2 \pi} \right) d {\tau} \ d T  \\
\nonumber
&&\hspace{-0.2in} \times e^{ i \frac{M}{H} \left(  {\bar \omega} U + \Omega u - \Omega {\bar u}_c \right) }  \left( {\bar \omega}^2- {k}_{12}^2 \right)^{-1/2} a({\tau} + T/2)^2 a({\tau} -T/2)^2 \\
\label{sma2}
&& \times {\bar \varphi}_{{\bf k}_1}  {\Phi}_{{\bf k}_2} ({\tau}+T/2) {\bar \varphi}^*_{{\bf k}_3}  {\Phi}^*_{{\bf k}_4} ({\tau}-T/2) .
\end{eqnarray}
The integral over $\Omega$ then constrains the average time $\tau$ to be near the stationary phase ${\bar \tau}_c$,
\begin{eqnarray}
\label{omegaint}
&& \hspace{-0.2in} \int_{0}^{ {\bar \omega}-k_{12}} d \Omega \ e^{-i \frac{M}{H} \Omega \left( {\bar u}_c - u \right) } \\
\nonumber
 &\approx&  ({\bar \omega}-k_{12} ) \exp - \frac{1}{3!} \left[ \frac{M}{H} ({\bar \omega}-k_{12} ) \left( {\bar u}_c - u \right) \right]^2 \\
 \nonumber
  && \hspace{-0.3in} + \frac{iM}{2H} ({\bar \omega}-k_{12} )^2 \left( {\bar u}_c - u \right) \exp - \frac{1}{12} \left[ \frac{M}{H} ({\bar \omega}-k_{12} ) \left( {\bar u}_c - u \right) \right]^2 .
 \end{eqnarray}
Summing the two integration regions in (\ref{sma2}) means only the real part is kept for diagram A.  Finally, consider the integral over ${\bar \omega}$: 
\begin{eqnarray*}
I_1(k_{12},u,U) \equiv \\
&& \hspace{-0.6in} \int_{k_{12}}^\infty \frac{  d {\bar \omega}({\bar \omega} - k_{12})}{  \sqrt{ {\bar \omega}^2 -k_{12}^2 }  }  e^{ - \frac{1}{3!} \left[ \frac{M}{H} ({\bar \omega}-k_{12} ) \left( {\bar u}_c - u \right) \right]^2 + i \frac{M}{H} U {\bar \omega}}.
\end{eqnarray*}
The steep exponential decay ensures that the bulk of the contribution is either in the neighborhoods of ${\bar \omega} = k_{12}$ or ${\bar u}_c({\bar \omega}) =u$.  In the former case, the leading behavior is the linear exponential $\exp - \left( \frac{M}{H} \right)^2 \epsilon$, giving a negligible total result.  The latter gives a much softer Gaussian $\exp - \left( \frac{M}{H} \right)^2 \epsilon^2$.  Thus we choose the latter, and approximate ${\bar \omega} \approx \sqrt{ k_{12}^2 + u^{-2}} + \epsilon$ for small $\epsilon$ which we formally extend over the whole axis:
\begin{eqnarray*}
 & & \hspace{-0.1in} I_1 \approx e^{i \frac{M}{H} U \sqrt{ k_{12}^2 + u^{-2}}}  \left(\sqrt{ k_{12}^2 u^2 + 1} - k_{12} |u| \right) \times \\
 && \hspace{-0.1in} \int _{-\infty} ^\infty d \epsilon \ e^{ -\frac{1}{3!} \left[ \frac{M}{H} \left(\sqrt{ k_{12}^2 u^2 + 1} - k_{12} |u|\right) \sqrt{ k_{12}^2 u^2+ 1} |u| \epsilon \right]^2 + i \frac{M}{H} U \epsilon} \\
&& \hspace{-0.2in} =  \frac{\sqrt{ 6 \pi } H}{M}  \frac{e^{i \frac{M}{H} U \sqrt{ k_{12}^2 + u^{-2}}}}{|u| \sqrt{k_{12}^2 u^2 + 1}}  e^{ - \frac{3U^2}{2 \left[ \left( \sqrt{ k_{12}^2 u^2+ 1} - k_{12} |u| \right) \sqrt{ k_{12}^2 u^2 + 1} |u| \right]^2 } }.
\end{eqnarray*}

The effective interaction for diagram A can then be written in the form
\begin{eqnarray*}
 \mathcal S^A_{\rm eff} &=& \int d \tau dT \ i \Gamma( { k}_{12}, \tau, T) a({\tau} + T/2)^2 a({\tau} -T/2)^2 \\
&& \hspace{0.4in} \times {\bar \varphi}_{{\bf k}_1}  {\Phi}_{{\bf k}_2} ({\tau}+T/2) {\bar \varphi}^*_{{\bf k}_3}  {\Phi}^*_{{\bf k}_4} ({\tau}-T/2) 
\end{eqnarray*}
where the effective quartic coupling is
\begin{equation}
\label{gamma}
 \Gamma( {k}_{12}, \tau, T) = \frac{g^2}{8 \pi HM} I_1 \left(k_{12}, \frac{\tau H}{M}, \frac{T H}{M} \right) . 
 \end{equation}
\begin{figure}
\begin{center}
\includegraphics[scale=0.7]{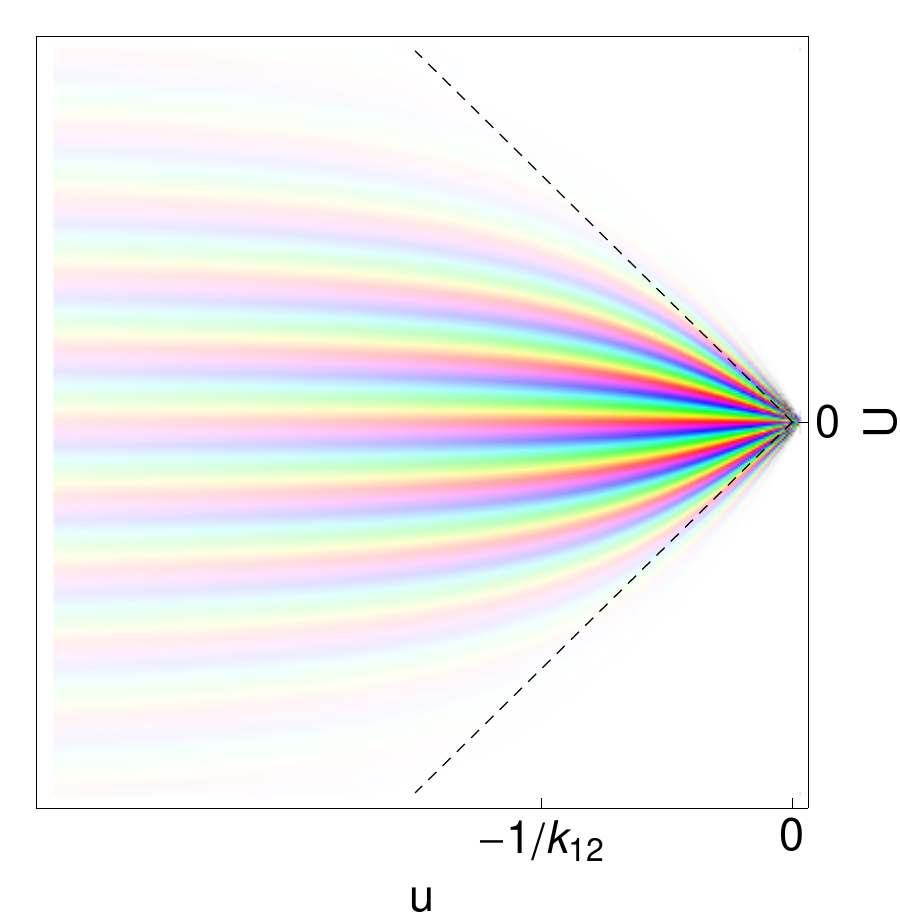}
\caption{Effective density matrix distribution $I_1$, with the phase represented by color and the magnitude represented by intensity.  The dashed lines denote $U = \pm |u|$.}
\label{density}
\end{center}
\end{figure}
This is the main result of this article.  The effective coupling (\ref{gamma}) is explicitly scale-invariant since the $1/\tau$ in $I_1$ will cancel the $dT$ in the measure.  Figure~\ref{density} shows the distribution as a function of $u$ and $U$, illustrating the delocalization, rapid phase oscillation, and behavior crossover when $|u| k_{12} \sim 1$.
\subsection{Diagrams B and C}
The action for diagrams B and C is similar to (\ref{sma}),
\begin{eqnarray*}
\mathcal S_{\rm eff}^{B,C} [ {\bar \varphi}, \Phi ] &=&\frac{g^2}{2i} \int d \tau_1\ a(\tau_1)^4 \ d \tau_2 \ a(\tau_2)^4 \\
\nonumber
&&  \hspace{-0.5in} \times {\bar \varphi}_{{\bf k}_1}  {\Phi}_{{\bf k}_2} (\tau_1) \theta(\tau_1-\tau_2) \left[ V_{{\bf k}_{12} } (\tau_1) V^*_{{\bf k}_{12} } (\tau_2) \right. \\
\nonumber
&& \hspace{-0.6in} \left.  - V^*_{{\bf k}_{12} } (\tau_1) V_{{\bf k}_{12} } (\tau_2) \right]  \left[ {\bar \varphi}^*_{{\bf k}_3}  {\bar \varphi}^*_{{\bf k}_4} ({\tau}_2) + \frac{1}{4}  {\Phi}^*_{{\bf k}_3}  {\Phi}^*_{{\bf k}_4} ({\tau}_2) \right].
\end{eqnarray*}
Upon Fourier expansion of the fields, the Heaviside function for the retarded Green's function enforces $|\omega_2|~<~|\omega_1|$.  The action then reduces to an answer identical to (\ref{sma2}) but with a different measure for $\Omega$,
\begin{eqnarray*}
\nonumber
&& \hspace{-0.3in} \mathcal S_{\rm eff}^{B,C} = \frac{ \pi g^2}{2iHM} \int_{k_{12}}^\infty \frac{d {\bar \omega}}{2 \pi} \left( \int_0^{{\bar \omega}-k_{12}} \frac{d \Omega}{2 \pi} - \int_{-({\bar \omega}-k_{12})}^0 \frac{d \Omega}{2 \pi} \right) \ d {\tau} \ d T   \\
&& \hspace{-0.3in} \times e^{ i \frac{M}{H} \left(  {\bar \omega} U + \Omega u - \Omega {\bar u}_c \right) } \left( {\bar \omega}^2- {k}_{12}^2 \right)^{-1/2} a({\tau} + T/2)^2 a({\tau} -T/2)^2 \\
\nonumber
&& \hspace{-0.3in} \times {\bar \varphi}_{{\bf k}_1}  {\Phi}_{{\bf k}_2} ({\tau}+T/2) \left[ {\bar \varphi}^*_{{\bf k}_3}  {\bar \varphi}^*_{{\bf k}_4} ({\tau}-T/2) + \frac{1}{4}  {\Phi}^*_{{\bf k}_3}  {\Phi}^*_{{\bf k}_4} ({\tau}-T/2) \right] .
\end{eqnarray*}
Now, only the imaginary part of (\ref{omegaint}) is kept.  The integral over ${\bar \omega}$ can be defined analogously and is
\begin{eqnarray*}
&& \hspace{0.1in} I_2(k_{12},u,U) \approx \frac{M}{2H} e^{i \frac{M}{H} U \sqrt{ k_{12}^2 + u^{-2}}} \times \\
 && \hspace{-0.1in}  \left(\sqrt{ k_{12}^2 u^2 + 1} - k_{12} |u| \right)^2    \sqrt{ k_{12}^2 u^2 + 1} |u| \times \\
 && \hspace{-0.1in} \int _{-\infty} ^\infty d \epsilon\ \epsilon \ e^{ -\frac{1}{12} \left[ \frac{M}{H} \left(\sqrt{ k_{12}^2 u^2 + 1} - k_{12} |u|\right) \sqrt{ k_{12}^2 u^2+ 1} |u| \epsilon \right]^2 + i \frac{M}{H} U \epsilon} \\
&& \hspace{-0.1in} \approx \sqrt{ 3 \pi } e^{i \frac{M}{H} U \sqrt{ k_{12}^2 + u^{-2}}} \left(\sqrt{ k_{12}^2 u^2 + 1} - k_{12} |u| \right) \times \\
&&  \sqrt{ k_{12}^2 + u^{-2}} \ e^{ - \frac{3U^2}{ \left[ \left( \sqrt{ k_{12}^2 u^2+ 1} - k_{12} |u| \right) \sqrt{ k_{12}^2 u^2 + 1} |u| \right]^2 } }.
\end{eqnarray*}
The final answer is then
\begin{eqnarray*}
 \mathcal S_{\rm eff}^{B,C} &=& \int d \tau dT \ \lambda( { k}_{12}, \tau, T) a({\tau} + T/2)^2 a({\tau} -T/2)^2 \\
&& \hspace{0.2in} \times {\bar \varphi}_{{\bf k}_1}  {\Phi}_{{\bf k}_2} ({\tau}+T/2) \Big[ {\bar \varphi}^*_{{\bf k}_3}  {\bar \varphi}^*_{{\bf k}_4} ({\tau}-T/2) \\
&&  \hspace{0.8in} + \frac{1}{4}  {\Phi}^*_{{\bf k}_3}  {\Phi}^*_{{\bf k}_4} ({\tau}-T/2) \Big].
\end{eqnarray*}
where
\begin{equation}
\label{lambda}
 \lambda( {k}_{12}, \tau, T) = \frac{g^2}{4 \pi HM}  I_2 \left(k_{12}, \frac{\tau H}{M}, \frac{T H}{M} \right)  , \hspace{0.2in} T \geq 0.
\end{equation}
This distribution is similar to (\ref{gamma}) but with a steeper Gaussian damping.  
\subsection{Validity as an Effective Action and \\ Higher-Order Corrections}
As a check that these effective interactions faithfully represent the original action (\ref{sorig}), let us (schematically) consider the  late-time $\varphi^4$-correlation function as computed in \cite{Jackson:2012fu},
\begin{eqnarray}
\label{phi4corr}
 \langle { \varphi}^4 \rangle &\equiv& \langle {\rm in} | e^{i \int_{- \infty}^0 d \tau' H} {\bar \varphi}(0)^4 e^{-i \int_{- \infty}^0 d \tau' H} | {\rm in} \rangle \\
 \nonumber
 && \hspace{-0.5in} = \langle {\rm in} | \int d \tau d T \ {\bar \varphi}(0)^4  {\bar \varphi} \Phi(\tau+T/2) \Big[ \Gamma(k_{12},\tau,T) {\bar \varphi} \Phi(\tau-T/2)  \\
  \nonumber
 && \hspace{-0.5in} + \lambda (k_{12},\tau,T) \left(  {\bar \varphi} ^2(\tau-T/2) +  \frac{1}{4} \Phi^2(\tau-T/2)  \right) \Big] | {\rm in} \rangle. 
 \end{eqnarray}
Contraction replaces the internal fields as $\varphi _{\bf k}(\tau)~\sim~e^{-i k \tau}$, and so the correlation function contains the oscillatory term
\begin{equation}
\label{effphase}
 e^{-i \frac{M}{H} \left[ (k_1 + k_2) (u+U/2) - (k_3 + k_4)(u-U/2) - U \sqrt{ k_{12}^2 + u^{-2}} \right]} . 
 \end{equation}
Extremizing this phase with respect to $u$ and $U$ gives the solutions
\begin{eqnarray}
\label{ueff}
u^{-2}_c &=&  \frac{1}{4} \left( k_1 + k_2 + k_3 + k_4 \right)^2 - k_{12}^2, \\
\nonumber
U_c &=& \frac{1}{2} \left[ \left( k_1 + k_2 \right)^2 - \left( k_3 + k_4 \right)^2 \right] |u_c|^{3}.
\end{eqnarray}
The energy transferred in the effective interaction is
\begin{eqnarray}
\nonumber
 \frac{\Delta E}{M} &=& (k_1 + k_2) \left( u_c+\frac{U_c}{2} \right) - (k_3 + k_4) \left( u_c -\frac{U_c}{2} \right) \\
 \label{etransfer}
 &=& - \frac{\left[ (k_1+k_2) - (k_3+k_4) \right] k_{12}^2}{\left[ \frac{1}{4} \left( k_1 + k_2 + k_3 + k_4 \right)^2 - k_{12}^2 \right]^{3/2}}.
\end{eqnarray}
As long as this is small, the effective action is trustworthy. One easily finds that this can be written as
\begin{equation}
\label{deltaeu}
 \frac{\Delta E}{M} = - \frac{U_c u_c^{-2}}{\sqrt{k^2_{12} + u_c^{-2}}}
\end{equation}
explaining why the last term in (\ref{effphase}) is simply the approximation to the phase integral in (\ref{vdef}), 
\[ \int_{u-U/2}^{u+U/2} du' \sqrt{k^2_{12} + u^{'-2}} \approx U \sqrt{k^2_{12} + u^{-2}} + \mathcal O(U^3).  \]
Expansion around the stationary phase solutions (\ref{ueff}) then gives the low-energy transfer approximation to the correlation (\ref{phi4corr}).

Suppose that we now return to (\ref{sma0}) and expand to first nontrivial order in $\Omega$,
\[ \hspace{-0.1in} \mathcal C_1\mathcal C^*_1 \approx  \frac{\pi | {\bar u}_c|}{HM} \left[ 1+ \frac{ 1}{8} |u_c|^4({\bar \omega}^2+k_{12}^2) \Omega^2 \right] e^{-i \frac{M}{H} \Omega {\bar u}_c } . \]
This merely modifies (\ref{gamma}) as
\[ \Gamma = \frac{g^2}{8 \pi HM} \left[ 1 +  \frac{U^2 u^{-2} (2 k_{12}^2 + u^{-2})}{8(k^2_{12} + u^{-2})}  \right] I_1 \]
and similarly for $\lambda$.  Via (\ref{deltaeu}) we see this is simply the $(\Delta E/M)^2$ correction to the original result.  This natural expansion of the coupling in terms of the energy transfer $\Delta E / M$ is the de Sitter analogue of (\ref{gfnaive}).  One may then constrain high-energy models by comparing the theoretically-obtained coefficients $\Gamma_i$ in the series
\[ \Gamma = \Gamma_0 + \Gamma_1 \frac{\Delta E}{M} + \frac{1}{2} \Gamma_2 \left(  \frac{\Delta E}{M} \right)^2 + \cdots \]
to observation via (\ref{etransfer}).  This is not quite true for $\lambda$, however, since its contribution in correlation functions will cancel to leading order in $H/M$ and hence would remain unconstrained \cite{Jackson:2010cw, Jackson:2011qg, Jackson:2012fu}.

\newpage
\subsection{Discussion and Conclusion}
In this article we have computed the effective interaction of a heavy field in a de Sitter background.  The effective de Sitter couplings (\ref{gamma}) and (\ref{lambda}) underscore how misleading the flat space approximation (\ref{gfnaive}) is when constructing inflationary models.  Due to the rapidly varying phase factor, the variance in interaction-time-difference is always negative, $\langle T^2 \rangle < 0$, and we cannot even in principle approximate these as time-localized interactions.  Attempting to do so quickly fails,
\[ \Gamma(k_{12},\tau,T) \approx \Gamma(k_{12},\tau,0)+ T \frac{\partial \Gamma(k_{12},\tau,0)}{\partial T} \sim \frac{M}{H} . \]

The overall magnitude of the effective couplings, $H/M$, was anticipated  as coming from modifications to the vacuum state encoded in the density matrix \cite{Schalm:2004qk,Easther:2001fi, Easther:2001fz, Easther:2002xe}.  This can now be understood as a simple consequence of the natural rescaling of time by $H/M$, and the fact that interactions are canonically normalized as $\int d \tau \lambda_n \varphi^n$.  These results also make clear that the density matrix will generically be corrected in cosmological effective field theory.  It will be important to integrate these results with previous work done on background evolution \cite{Achucarro:2012sm, Cespedes:2012hu, Achucarro:2010da, Achucarro:2010jv} and parameterizing possible effective action terms \cite{Senatore:2010wk, Weinberg:2005vy, Weinberg:2006ac, Weinberg:2008hq, Weinberg:2010wq}.

The simple form of the distribution suggests that the Gaussian-with-rapid-oscillations distribution may be universal for all effective interactions in de Sitter space, in the same way that systems in equilibrium universally tend towards the thermal distribution $P \sim e^{-TE}$.  It would be interesting to understand the theoretical basis for this.  

{\bf Acknowledgments:}
I would like to thank A.~Ach\'{u}carro, F.~Bouchet, B.~Greene, D.~Langlois, G.~Palma, M.~Sakellariadou, K.~Schalm, D.~Steer and B.~Wandelt for discussions and collaborations.


\begin{thebibliography}{99}

\small
\parskip=0pt plus 2pt

\bibitem{Birrell:1982ix}
  N.~D.~Birrell and P.~C.~W.~Davies,
{\it  Cambridge, UK: Univ. Pr. (1982) 340p}

\bibitem{planck}
    [Planck Collaboration],
  ``Planck: The scientific programme,''
  arXiv:astro-ph/0604069.

\bibitem{Guth:1980zm}
  A.~H.~Guth,
  Phys.\ Rev.\  D {\bf 23}, 347 (1981).
  
\bibitem{Linde:1981mu}
  A.~D.~Linde,
  Phys.\ Lett.\  B {\bf 108}, 389 (1982).

\bibitem{Albrecht:1982wi}
  A.~Albrecht and P.~J.~Steinhardt,
  Phys.\ Rev.\ Lett.\  {\bf 48}, 1220 (1982).

\bibitem{Linde:1983gd}
  A.~D.~Linde,
  Phys.\ Lett.\  B {\bf 129}, 177 (1983).
  
\bibitem{Jackson:2010cw}
  M.~G.~Jackson and K.~Schalm,
Phys. \ Rev. \ Lett. {\bf 108}, 111301 (2012)
  arXiv:1007.0185 [hep-th].
  
\bibitem{Jackson:2011qg}
  M.~G.~Jackson and K.~Schalm,
  arXiv:1104.0887 [hep-th].
  
\bibitem{Jackson:2012fu} 
  M.~G.~Jackson and K.~Schalm,
  arXiv:1202.0604 [hep-th].
  
\bibitem{Brandenberger:1999sw}
  R.~H.~Brandenberger,
  arXiv:hep-ph/9910410.
  
\bibitem{Schalm:2004qk} 
  K.~Schalm, G.~Shiu and J.~P.~van der Schaar,
  JHEP {\bf 0404}, 076 (2004)
  [hep-th/0401164].
  
  \bibitem{Easther:2001fi}
  R.~Easther, B.~R.~Greene, W.~H.~Kinney and G.~Shiu,
  Phys.\ Rev.\  D {\bf 64}, 103502 (2001)
 [arXiv:hep-th/0104102].


\bibitem{Easther:2001fz}
  R.~Easther, B.~R.~Greene, W.~H.~Kinney and G.~Shiu,
  Phys.\ Rev.\  D {\bf 67}, 063508 (2003)
 [arXiv:hep-th/0110226].

\bibitem{Easther:2002xe}
  R.~Easther, B.~R.~Greene, W.~H.~Kinney and G.~Shiu,
  Phys.\ Rev.\  D {\bf 66}, 023518 (2002)
  [arXiv:hep-th/0204129].


\bibitem{Kaloper:2002uj}
  N.~Kaloper, M.~Kleban, A.~E.~Lawrence and S.~Shenker,
  Phys.\ Rev.\  D {\bf 66}, 123510 (2002)
  [arXiv:hep-th/0201158].


\bibitem{Kaloper:2002cs}
  N.~Kaloper, M.~Kleban, A.~Lawrence, S.~Shenker and L.~Susskind,
  JHEP {\bf 0211}, 037 (2002)
  [arXiv:hep-th/0209231].


\bibitem{Burgess:2002ub}
  C.~P.~Burgess, J.~M.~Cline, F.~Lemieux and R.~Holman,
  JHEP {\bf 0302}, 048 (2003)
  [arXiv:hep-th/0210233].


\bibitem{Bunch:1978yq}
  T.~S.~Bunch and P.~C.~W.~Davies,
  Proc.\ Roy.\ Soc.\ Lond.\  A {\bf 360}, 117 (1978).

\bibitem{Calzetta:1986cq}
  E.~Calzetta and B.~L.~Hu,
  Phys.\ Rev.\  D {\bf 37}, 2878 (1988).

\bibitem{Achucarro:2012sm} 
  A.~Achucarro, J.~-O.~Gong, S.~Hardeman, G.~A.~Palma and S.~P.~Patil,
  arXiv:1201.6342 [hep-th].
  
\bibitem{Cespedes:2012hu} 
  S.~Cespedes, V.~Atal and G.~A.~Palma,
  arXiv:1201.4848 [hep-th].
  
\bibitem{Achucarro:2010da} 
  A.~Achucarro, J.~-O.~Gong, S.~Hardeman, G.~A.~Palma and S.~P.~Patil,
  JCAP {\bf 1101}, 030 (2011)
  [arXiv:1010.3693 [hep-ph]].
  
\bibitem{Achucarro:2010jv} 
  A.~Achucarro, J.~-O.~Gong, S.~Hardeman, G.~A.~Palma and S.~P.~Patil,
  Phys.\ Rev.\ D {\bf 84}, 043502 (2011)
  [arXiv:1005.3848 [hep-th]].

\bibitem{Senatore:2010wk} 
  L.~Senatore and M.~Zaldarriaga,
  arXiv:1009.2093 [hep-th].
  
  \bibitem{Weinberg:2005vy}
  S.~Weinberg,
  Phys.\ Rev.\  D {\bf 72}, 043514 (2005)
  [arXiv:hep-th/0506236].

\bibitem{Weinberg:2006ac}
  S.~Weinberg,
  Phys.\ Rev.\  D {\bf 74}, 023508 (2006)
  [arXiv:hep-th/0605244].

\bibitem{Weinberg:2008hq}
  S.~Weinberg,
  Phys.\ Rev.\  D {\bf 77}, 123541 (2008)
  [arXiv:0804.4291 [hep-th]].

\bibitem{Weinberg:2010wq}
  S.~Weinberg,
  Phys.\ Rev.\  D {\bf 83}, 063508 (2011)
  [arXiv:1011.1630 [hep-th]].


%
\end{thebibliography}
\end{document}